# On the structurization of coal dust precipitations and their influence on aerodynamic resistance by granulated mediums in air filters at nuclear power plants

I.M. Neklyudov, O.P. Ledenyov, L.I. Fedorova, P.Ya. Poltinin

*National Scientific Centre Kharkov Institute of Physics and Technology, Academicheskaya 1, Kharkov 61108, Ukraine.*

The processes of structurization of dust precipitations in granulated filtering mediums, formed by the monolithic glass spherical granules with the diameters of 2*mm* and 3*mm,* are researched. The distinctions between the distributions of filtered coal dust masses in the air filters with cylindrical granules and the air filters with spherical granules, are found. The influences by the filtered dust masses on the air resistance of both the air filters with the cylindrical granules and the air filters with the spherical granules are described. The conclusions on a possibility of the use of various chemical adsorbents with different geometric forms and volumetric dimensions to improve the filtering properties of granulated filtering mediums in air filters at nuclear power plants are formulated.

## Introduction

The weighted dust masses in gas medium (aerosols) and in vacuum have a number of unique physical properties, which represent a considerable research interest. The weighted dust masses can create some specific structures, for example, the dust crystals, vortices and other formations, appearing at application of external electric fields (the dust plasma) as well as at influence by external forces and conditions, impacting the interaction between the weighted dust masses and the surrounding mediums during the dust transport processes [1, 2]. The air filters with the chemical granules with different geometric forms and volumetric dimensions, have these granulated filtering mediums.

The research on the physical features of formation of coal dust structures in the vertical air filters, filled by the coal granules, was conducted during the investigation of the influence by the fractional composition of air transferred coal dust on the increase of aerodynamic resistance by the *Iodine* absorbers in air ventilation systems at nuclear power plants (*NPP*) [3, 4]. The design of researched air filters [3, 4] was similar to the design of air filters with the filling pouring structure with the cylindrical granules of coal absorbent of type of *СКТ-3* (diameter 1,8 *mm*, length 3,2 *mm*) at *NPP*. Thus, the complex filtering system with the cylindrical coal granules and chaotic air channels was created in air filters with the purpose of air-dust mixture filtering during the air-dust aerosol pumping. The porosity of filling layer with absorbent granules (the relation between the total volume of empty cavities and the total volume of empty space and granules) was 0,23; the density of cylindrical coal granules packing was $\eta = 0,77$. The magnitude of aerodynamic resistance by the air filters increased during the coal dust fractions accumulation, which had place due to the nature of the structurization of coal dust precipitations and it's impact on the filling of empty cavities and constrictions, appearing in the granulated filtering medium at non-homogenous packing of cylindrical coal granules in an air filter. In [3], it was established that, at critical concentration of large coal dust particles (the particle's dimension of 10 *μm* and below) in the narrow surface layer of granulated filtering medium in an air filter, the monolithic layer consisting of big coal dust fractions can be created, blocking the penetration of air through an air filter. In this case, the aerodynamic resistance sharply increased in more than ten times, resulting in a malfunction by the air filters. In [4], when the fractional consistence of coal dust was biased to the region of small particles dimensions (the particle's dimension of 1 *μm* an below), the character of distribution of the coal dust particles along the length of an air filter was significantly changed. At the same coal dust loads, the surface layer of granulated filtering medium was transparent for the aerosol stream with the air-dust mixtures. During the initial process of small dust fraction accumulation, the magnitude of aerodynamic resistance by an air filter slowly increased in agreement with the linear law approximately. At the final phase of experiment, when the total mass share of small coal dust fraction was 9,3 %, the magnitude of aerodynamic resistance by an air filter increased in the two times only.

Comparing the obtained research results in [3] and [4], it is possible to make a conclusion that, in the case of big- and small- dust coal fractions, the observed features in the character of distribution of coal dust particles along the layer of granules have place because of a number of reasons. Among them, the relation between the dimensions of coal dust particles, transporting through an air filter, and the dimensions of empty cavities and transitional constrictions, created in the filling filtering layer in an air filter, can play an important role.

The main purpose of research is to find out an influence by the geometric form and dimensions of granules on the character of distribution of coal dust precipitations along the length of filling filtering layer, and as a result on the aerodynamic resistance by an air filter. At the change of geometric form of coal granules from the cylindrical shape to the spherical form, the variation of dimensions of empty cavities and constrictions, which serve to transport the aerosol with the air-dust mixture in a filling filtering medium in an air filter, takes place. The characteristic dimension of coal granules is an important parameter. Starting from some critical dimension of coal granules, the further decrease of granule's dimension results in an intensive accumulation of coal dust precipitations between the coal granules, because of the contraction of air channels in filling filtering medium in an air filter.

In this research, the modeling experiments with the use of a model of an air filter with the monolithic glass spherical granules with the diameter of 2 *mm* and 3 *mm*, re conducted. The density of glass is 2,461 *g/cm³*. Let us highlight that these spherical granules, because of their monoliticity, had no volumetric absorption properties. However, the main research subject is focused on the problems of structurization of coal dust precipitations rather than on the processes of chemical compounds absorption, hence such an interchange by the research objects in filling granulated medium is feasible. Let us note that the monolithic glass spherical granules had a good adhesion property in relation to the coal dust fraction. It appeared that the small coal dust fraction was able to cover the surface of monolithic glass spherical granules easily. Also, the monolithic glass spherical granules were able to capture the small coal dust fraction, keeping it on the surface, as in the case of cylindrical coal granules of the type of *СКТ-3*.

From the geometric point of view, the filling filtering medium, created by the granules with the spherical form, is most symmetrical, hence it is relatively simple for the research analysis. The presence of alternation between the empty enlargements (cavities) and the empty constrictions (draughts) is a characteristic property by the filling porous mediums. In the case of the spherical granules, the geometric form and dimensions of empty cavities and constrictions depend on the packing density of spherical granules, which create the porous structure in the filling filtering medium in an air filter [5].

In the research by V.P. Voloshin et al. [6], the models of the dense unordered packs of the same spherical granules with the diameter of *d* were created with the use of direct numerical computer modeling. The purpose of structural research is the understanding of the geometric structure of empty space between the spherical granules in the filling filtering medium in an air filter. The authors are interested in a wide scope of possible applications, based on the empty space geometric structure research results for the finding of appropriate solution of the problem of mass transfer in the porous medium, created by the spherical granules. At the dense packing of the spherical granules in the porous medium, the spherical granules are distributed homogeneously. The certain part of tetrahedral and octahedral configurations by the spherical granules are always present, however they don't create the crystal order in the geometric structure in the general case. In [6], the quantitative analysis of geometric structure of empty space in these models was conducted and their geometric structure was researched in details. The dense model with the spherical granules has the value of packing density (the degree of space filling) $\eta = 0,59$, that is slightly smaller than the value of packing density in the models with the cylindrical granules. The computer modeling of geometric structure provided the results on the geometric characteristics such as the volume of empty space between the spherical granules, radius of curvature of the empty enlargements (cavities) and the empty constrictions (draughts). In the case of the empty narrow constrictions (draughts), the radius is $R_0 = 0,081\ d$.

The distribution of radiuses of spheres *R*, written inside the empty cavities, has the maximums with the radius $R_1 = 0,12\ d$, which is approximately equal to the radius of the sphere, written inside the tetrahedral configuration of solid spheres, and $R_2 = 0,2\ d$, which is approximately equal to the radius of sphere, written inside the right octahedral configuration of solid spheres [5, 6].

In [6], the described dependences of the dimensions of constrictions and cavities on the diameter of the spherical granules, obtained in the completed modeling experiments with the use of a model of an air filter, allow to make an evaluation of the relations between the dimensions of coal dust particles, transferring in an air filter, and the dimensions of the constrictions and cavities in a filling filtering layer, consisting of the spherical granules. The influence by the various regimes of air dust mixtures aerosol transport on the coal dust precipitations structure creation can also be clarified. Let us note that the research on the similar, but less complex problem, related to the research on the features of gas transport at an absence of dust fraction, has an important impact on the design of nuclear reactors with the spherical fuel elements [7]. In this case, the variants of changes of cross-sections of the input and output gas collectors in the conditions of incoming and outgoing gases in the granulated medium, consisted of the spheres with the diameter of $d = 3...5\ mm$ were researched with the purpose of optimal selection of conditions of gas spread [8].

In this research paper, the research on the characteristic dependence of distribution of coal dust precipitations along the height of filling filtering medium with the spherical granules of different diameters is completed, using the air-blasting with the air-dust mixture with the small coal dust fractions with the particles diameter of 1 *μm* and smaller in the air filters at *NPP*. The influence by the coal dust precipitations distribution on the aerodynamic resistance of the air filters with the filling filtering medium, consisting of the spherical granules of different diameters, is researched in a series of modeling experiments.

## Methods of experiment

The research was conducted with the application of a model of a vertical air filter with the 10 times smaller diameter, comparing to a real air filter, as shown in Fig. 1. The height of filling filtering medium with the mono-lithic glass spherical granules in a researched model of a vertical air filter is 30 *cm*, which is same, comparing to a real air filter at *NPP*. The magnitudes of aerodynamic resistances in the both air filters were same, because of the chosen parameters. The magnitude of aerodynamic resistance was measured by the water manometer as the difference between the input air pressure and the output air pressure *ΔP* in an air filter. The ten metallic containers with the monolithic glass spherical granules, which were placed between the two grids with the cells of smaller diameter than the granules diameter, were loaded. The container, which is a source of coal dust, was situated above an air filter. This container was filled by the mixture, containing the coal granules and the coal dust (the mass share of dust in relation to the total mass of granules was 1,5% in a single unit of coal dust source in a container). With the aim of simplification of comparison of the present research results with the results by other researchers in [3, 4, 9], all the obtained data on the relative mass of dust content are normalized to the density of coal granules. The content of a container, which played a role of dust source, was renewed before each cycle in the experiment. At the application of air stream to an air filter, the incoming dust was partly accumulated and partly jettisoned by an air filter, leaving it with the outgoing air. The mass of accumulated dust in an air filter was calculated as a difference between the mass of a container before the experiment and the mass of a container after the experiment. The model of an air filter, which was developed by the authors, allows to obtain the research data on the distribution of dust along the height of granules layer in the filling filtering medium after the completion of experiment. It can be done by the separation of weight of coal dust precipitations accumulated in the containers, and by the measurement of their total mass. The following variables are used: $M_0$ is the total mass of granules in an air filter; $M_i$ the mass of granules in each container ($i$ − from 1 to 10); $m_0$ is the total mass of dust, captured by an air filter during every regular experiment; $m_i$ is the mass of dust, precipitated in each container to the end of research; $h$ is the distance from the surface of an air filter. In the experiments, the monolithic glass spherical granules (balls) were used.

The two experiments were conducted: 1) with the monolithic glass spherical granules with the diameter of 2 *mm*; 2) with the monolithic glass spherical granules with the diameter of 3 *mm*. Let us note that the volume of a cylindrical coal granule in the case of the adsorbent of the type of *CKT-3* in [3, 4, 9] is equal to the volume of a conditional spherical granule with the diameter $d≈2,5mm$. In distinction from the cylindrical coal granules in the adsorbent of the type of *CKT-3,* in the completed experiments with the models of air filters with the monolithic glass spherical granules, it was established that the difference between the diameters of monolithic glass spherical granules on the order of magnitude of ±0,5*mm* is enough to have a great impact on the properties of coal dust precipitations structure creation, and hence on the magnitude of aerodynamic resistance by the air filters.

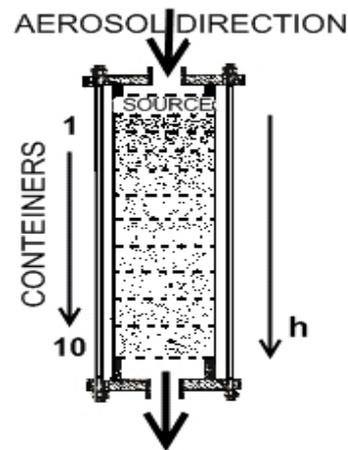

Fig. 1. Scheme of a vertical air filter.

Let us note that the filling filtering medium with the spherical granules was exposed to the strong vibration with the purpose to improve the packing density of spherical granules in a model of a vertical air filter. The value of porosity of filling filtering medium was measured as a relation between the volume of empty cavities and constrictions around the monolithic glass spherical granules to the total volume of filling filtering medium (monolithic glass spherical granules + empty space). The value of porosity of filling filtering medium with the monolithic glass spherical granules with the diameter of 2 *mm* was 0,35; and the value of porosity of filling filtering medium with the monolithic glass spherical granules with the diameter of 3 *mm* was 0,38. In [3, 4, 9], the value of porosity of filling filtering medium with the cylindrical coal granules was equal to 0,23.

## Measurements results: distribution of small dust coal fraction along vertical filter

In Fig. 2, in Curves 1, 2, the final distributions of mass share of small coal dust $m_i/(M_i+m_i)$ on the distance $h$ from the surface of filling filtering medium with the monolithic glass granules are shown in the researched cases of a model of an air filter with the filing filtering mediums with the monolithic glass spherical granules with the diameters of 2 *mm* and 3 *mm*. In Fig. 2a, in Curves 3, 4, the early researched dependences on the distributions of mass share of small- and big- coal dust $m_i/(M_i+m_i)$ on the distance from the surface $h$ of filling filtering medium in an air filter with the absorbent of the type of *CKT-3* are presented with the purpose of comparative analysis [3, 4]. As it is shown, the maximum of density of precipitated small coal dust fraction ~ *23 %* is observed in the narrow layer of 2 *cm* in close proximity to the surface of the filling filtering medium with the monolithic glass granules with the diameter of 2 *mm*. This maximum of density of precipitated small coal dust has similar characteristics, comparing to the maximum of density of precipitated big coal dust, obtained in the case of use of an air filter with the filling filtering medium with the absorbent of the type of *CKT-3* [3]. In Fig. 2, in Curve 1, the registered dependence with the sharp decrease (in 12 times) of mass share of coal dust from 23% to 2% on the distance from the surface $h=3cm$ of the filing filtering medium in an air filter is similar to the early observed dependences of density of small coal dust $m_i/(M_i+m_i)$ on the distance from the surface $h$ of the filing filtering medium in an air filter, and it can be described by the formula in [9]

$$C(z) = C_{Z=0}[1 - erfz] = C_{Z=0}erfcz, \quad (1)$$

where $erfz$ is the «integer of errors» by Gauss; $z = (3)^{1/2}(h-1)/2(Dt)^{1/2}$, where $D$ is the diffusion coefficient of dust mass, which can be determined from the experimental dependence in the researched case; $h$ is the distance from the surface; $t$ is the time. Further, the small packing is observed on the distance of $\sim 8$ $cm$ from the surface, then there is a fluent decrease of mass share of dust up to 0,2 % (in 100 times) on the distance of 13 $cm$ from the surface of the filing filtering medium in an air filter.

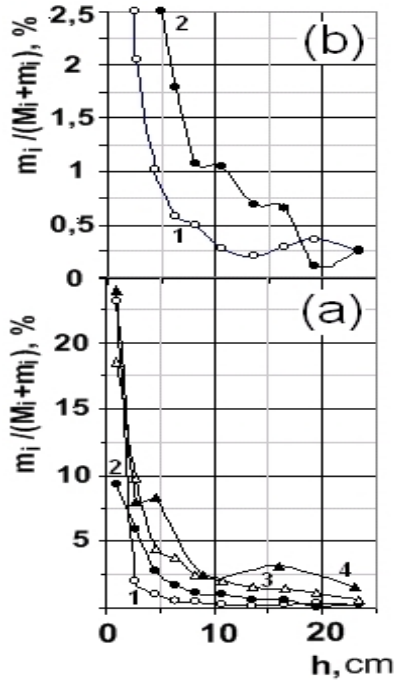

*Fig. 2. Distribution of mass share of coal dust fraction along vertical air filter at the end of series of experiments: 1, 2 – an air filter with monolithic glass spherical granules with diameters of 2 mm and 3 mm correspondingly; 3, 4 – an air filter with adsorbent of type of CKT-3. Dimensions of coal dust particles: 1, 2, 3 – smaller than 1 μm, 4 – smaller than 10 μm; i – from 1 to 10 μm.*

The maximum of mass share of coal dust up to 0,4% is registered on the distance of 20 $cm$ from the surface of filling filtering medium in an air filter. These maximums of mass shares of coal dust were described by the *Gauss distribution* formula in [9]

$$C(h) \propto \left(Q_0/(\pi D_i t)^{1/2}\right)\exp((h-h_0)^2/4D_i t) \quad (2)$$

The dependences (1) and (2) are shown in Fig. 3. The Curve 1 corresponds to the smooth part of coal dust masses distribution. In Curve 2, the maximums of coal dust accumulation are situated on the distances of 8 $cm$ and 20 $cm$ from the surface of filing filtering medium in an air filter. The Curve 3 (see Fig. 3) corresponds to the Curve 1 (see Fig. 2). The sub-surface layer in an air filter with the filling filtering medium with the monolithic glass spherical granules with the diameter of 3 $mm$ contains the coal dust mass share of $\sim 10$ % (see Fig 2, Curve 2). The coal dust mass share decreases in 10 times up to $\sim 1$ % on the distance of 8 $cm$ from the surface of filling filtering medium in an air filter. Then, there is a non-monotonic decrease of mass share of small coal dust $m_i/(M_i+m_i)$ below 0,1 % on the distance of 20 $cm$ from the surface of filling filtering medium in an air filter, which is accompanied by an appearance of the two maximums on the distances of 11 $cm$ and 17 $cm$ from the surface of filling filtering medium in an air filter.

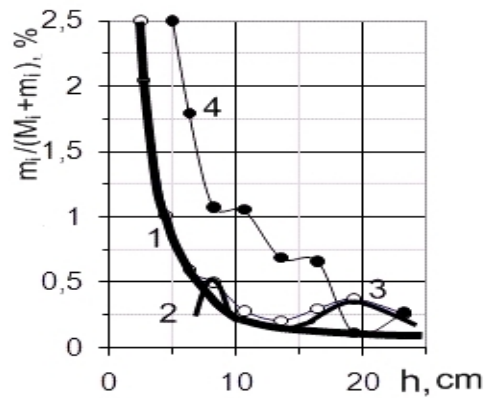

*Fig. 3. Distribution of mass share of coal dust fraction along vertical air filter as perfect dependence (1) and dependence with two maximums (2, 3) in the case of air filter with monolithic glass spherical granules with diameter of 2 mm. Curve (4) with black dots describes air filter with monolithic glass spherical granules with diameter of 3 mm.*

Practically, the small maximum of mass share of coal dust is observed on the distance of 25 $cm$ from the surface of filing filtering medium in an air filter. In Fig. 2, Curve 2, the dependence of mass share of small coal dust $m_i/(M_i+m_i)$ on the distance $h$ from the surface of filling filtering medium with the monolithic glass granules with the diameter of 3 $cm$ is similar to the dependence of mass share of small coal dust $m_i/(M_i+m_i)$ on the distance $h$ from the surface of filling filtering medium with the monolithic glass granules with the diameter of 2 $cm$, however the maximums of coal dust mass density are situated on the different characteristic distances and have different relative values of accumulated coal dust (see Fig. 3, Curve 4).

The mass of dust, jettisoned with air stream, was calculated as a difference between the maximum mass of dust, captured by an air filter, and the total mass of dust, precipitated in an air filter at the end of experiment. In the air filters with the spherical granules with the diameters of 2 $mm$ and 3 $mm$, the 20,1 % and 83,5 % of dust mass from the total dust mass were carried out by the air stream correspondingly.

The mass of dust, taken out by the air stream from the air filter of the model of *СКТ-3*, in proportion to the total mass, was [3, 4, 9]:1) Big dust fraction: 61 %; 2) Small dust fraction: 44 %.

### Influence by the geometric dimension of absorbing granules in filling filtering medium on the aerodynamic resistance by an air filter

During the experimental research, the dependence of aerodynamic resistance $\Delta P$ on the volumetric air stream $J$ was researched for an every increasing value of mass share of dust: $m_o/(M_o+m_o)$.

In Fig. 4, the selected research results, including the characteristic dependences $\Delta P(J)$ in the air filters with the monolithic glass spherical granules with the diameters of 2 *mm* and 3 *mm* as well as with the granules of absorbent of the type of *СКТ-3* at various values of mass share of introduced small coal dust are shown.

The charts, describing the transport of big coal dust fraction through the filing filtering medium with the absorbent of the type of *СКТ-3,* are also shown in Fig. 4. During the increase of effective mass of incoming coal dust in an air filter, the increase of aerodynamic resistance by an air filter with the filling filtering medium with the monolithic glass granules with the diameter of 2 *mm* significantly advances both the increase of aerodynamic resistance by an air filter with the filing filtering medium with the monolithic glass granules with the diameter of 3 *mm* (see the Curves 3 и 6) and the increase of aerodynamic resistance by an air filter with the filling filtering medium with the cylindrical coal granules (see the Curves 5 and **3,** 9 and **4**). The increase of aerodynamic resistance by an air filter with the filling filtering medium with the monolithic glass spherical granules with the diameters of 3 *mm* is significantly slow, comparing to the increase of aerodynamic resistance by an air filter with the filling filtering medium with the cylindrical coal granules (see the Curves 6 and **2).**

In Fig. 5, in Curves 1–4, the dependences of aerodynamic resistance $\Delta P^*$, normalized to the constant air stream $J^* = 15$ $m^3/hour$, on the relative mass share of introduced coal dust are shown. The early obtained empirical exponential dependence $\Delta P(J)$ [3] and the graphs in Fig. 4 were used during the calculations. The following experimental graphs are obtained in the case of air blow with the use of small coal dust fraction in application to the various filling filtering mediums in an air filter:

Curves 1 and 2 – the air filters are filled with the filling filtering medium with the monolithic glass granules with the diameters of 2 *mm* and 3 *mm* correspondingly and the small coal dust fraction is applied;

Curve 3 – the air filter is filled with the filling filtering medium with the granules of absorbent of the type of *СКТ-3* and the small coal dust fraction is applied;

Curve 4 – the air filter is filled with the filling filtering medium with the granules of absorbent of the type of *СКТ-3* and the big coal dust fraction is applied.

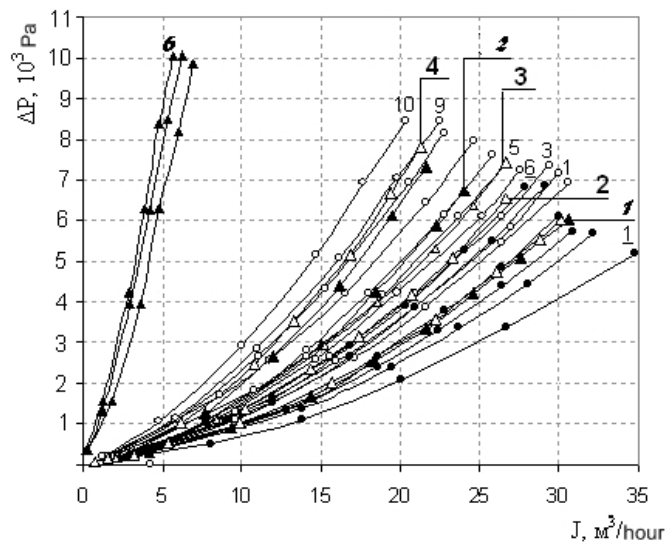

*Fig. 4. Dependence of aerodynamic resistance by vertical absorber on volume of air stream at certain mass share of dust fraction, captured by air filter, %;*

*Dimension of coal dust particles:○ – (1–10), ● - (1–6), △ - (1-4) – 1 μm and below; ▲ – (1-6) – 10 μm and below.
Air filter is filled with : 1) ○ – monolithic glass spherical granules with diameter of 2 mm:
1 (0), 2 (0,4), 3 (1,0), 4 (1,7), 5 (1,95), 6 (2,2), 7 (2,4), 8 (2,8), 9 (3,0),10 (3,1);
2) ● – monolithic glass spherical granules with diameter of 3 mm: 1 (0), 2 (3,4), 3 (4,5), 4 (6,7), 5 (7,6), 6 (10,5);
3) △ – granules of absorber of type of СКТ-3: 1 (0), 2 (1,4), 3 (4,3), 4 (9,3);
4) ▲ – granules of absorber of type of СКТ-3: 1 (0), 2 (5,9), 3 (6,7), 4 (8,8), 5 (9,1), 6 (9,2).*

In Fig. 5b, it is visible that, in the case without the dust load, the initial aerodynamic resistance by an air filter with the monolithic glass spherical granules with the diameter of 2 *mm*, is in 1,5 times higher than the initial aerodynamic resistance by an air filter with the monolithic glass spherical granules with the diameter of 3 *mm*. The aerodynamic resistance by an air filter with the monolithic glass spherical granules with the diameter of 2 *mm* is in 1,9 times higher than the maximal aerodynamic resistance by an air filter with the monolithic glass spherical granules with the diameter of 3 *mm* at $m_o/(M_o+m_o) = 10,5$ %, and it is in 1,2 times higher than the final aerodynamic resistance by an air filter with the filling filtering medium with the *CKT-3* absorber at the mass share of introduced coal dust precipitations in 9,3%. The aerodynamic resistance by an air filter with the monolithic glass spherical granules with the diameter of 3 *mm* increases much slowly than the aerodynamic resistance by an air filter with the filling filtering medium consisting of the coal granules. At the same mass share of the introduced coal dust precipitations in 9,3 *%*, the aerodynamic resistance by an air filter with the monolithic glass spherical granules with the diameter of 3 *mm* is in 1,7 times less than the aerodynamic resistance by an air filter with the filling filtering medium consisting of the coal granules. In an air filter with the monolithic glass spherical granules with the diameter of 3 *mm*, the maximum of aerodynamic resistance is 2850 *Pascale*, and it is observed at $m_o/(M_o+m_o) = 10,5$ %. The same magnitude of aerodynamic resistance in an air filter with the filling filtering medium consisting of the coal granules is reached, when the mass share of coal dust precipitations is 2,3 *%* only.

The comparative analysis of results (see Figs. 2 and 4) allows us to make the following conclusions. As it is shown in the graph 1 in Fig. 2a, in an air filter with the monolithic glass spherical granules with the diameter of 2 *mm*, the accumulated coal dust precipitations (around ~ 80 *%* of introduced mass) are mostly situated inside the narrow layer with the approximate thickness of 2 *cm* near the surface. The mass share of coal dust precipitations decreases in 100 *times* on the distance of 15 *cm* from the surface. At small increase of mass share of coal dust precipitations up to 3 %, the aerodynamic resistance increases sharply (Curve 1 in Fig. 5). This experimental result presents a direct evidence that there is a monolithic layer with the coal dust precipitations in close proximity to the surface of filling filtering medium in an air filter. This monolithic layer, composed of coal dust precipitations, can not be penetrated by the aerosol with the air-dust mixture.

It is necessary to note that, at final phase of experiment, the coal dust precipitations are accumulated above the surface of granulated filtering medium without the subsequent penetration inside an air filter.

The comparative analysis (see the Curves 2 and 3 in Fig. 5) shows that it is necessary to introduce up to the 7 *%* of mass share of coal dust precipitations in an air filter with the monolithic glass spherical granules with the diameter of 3 *mm* to reach the initial magnitude of aerodynamic resistance by an air filter, filled with the absorber of the type of *CKT-3*. As it was shown in the experimental research, the 80 *%* of introduced dust in an air filter, filled by the monolithic glass balls, outflows from an air filter together with the air stream. Therefore, at an increase of relative mass share of introduced dust up to 10,5 *%* in an air filter, the aerodynamic resistance by an air filter with the monolithic glass spherical granules with the diameter of 3 *mm* increases in 1,7 times only. The mass share of accumulated dust (16 *%*) non-monotonously decreases along the length of filling filtering medium (Curve 2 in Fig. 2), and can not serve as an obstacle to the dust fraction transport through the layers of monolithic glass spherical granules inside and to the aerosol's outflow by an air stream outside an air filter. However, in this case, the appearance of some maximums in the density of dust fraction distribution along an air filter is observed (Curve 2 in Fig. 2).

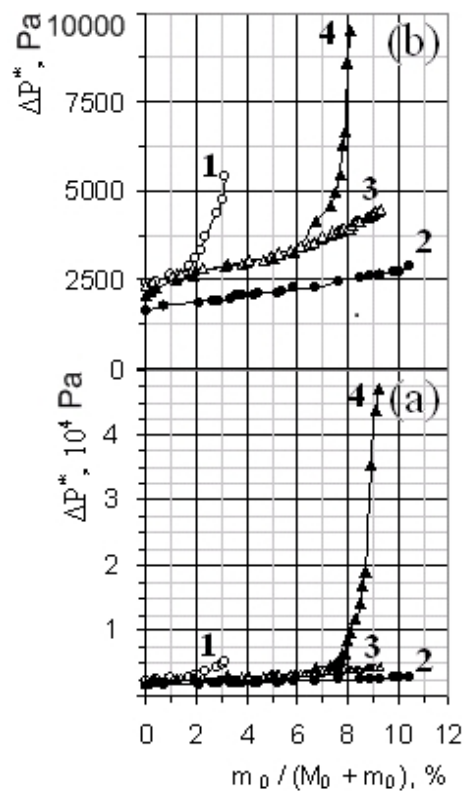

*Fig. 5. Dependence of aerodynamic resistance at air stream $J^*=15$ $m^3$/hour on relational mass share of dust fraction captured in air filter.*
*1, 2 - air filter is filled by monolithic glass spherical granules with diameter of: 1 – 2 mm, 2 – 3 mm;*
*3, 4 – air filter is filled by absorber of type of CKT-3.*
*Dimension of coal dust particles: 1, 2, 3 – 1 μm and below; 4 – 10 μm and below.*

## Conclusion

During the completed research, it is found that there is a significant difference in the characteristic distributions of small dust coal fractions along the height of air filters in the researched cases of spherical filtering coal granules with the diameters of 2 *mm* and 3 *mm* despite the fact that the values of porosity of filling filtering layers have 8% distinction only. The main influencing factor is a relation between the dimension of coal dust particles and the dimension of cavities and constrictions in the structure of porous filtering medium. In agreement with the calculations in [6], the dimension of cavities and draughts in the filling filtering layer with the spherical granules with the diameter of 2 *mm* is smaller in 1,5 times in comparison with the filling filtering layer with the spherical granules with the diameter of 3 *mm*. Correspondingly, in the case of the spherical granules with the diameter of 3 *mm*, the relation between the diameter of cavities (constrictions) and the dimension of coal dust particles will be in 1,5 times bigger, comparing to the case of the spherical granules with the diameter of 2 *mm*. At the equal coal dust loads, the physical mechanisms of interaction between the filtering medium and the transported coal dust have significant differences in the researched cases of various air filters:
1) The coal dust with small fractions is freely transported though the structure of empty cavities in the filling filtering medium with the spherical granules with the diameter of 3 *mm* in an air filter (the air filter jettisons up to 80 % of the coal dust mixed with the air).
2) The coal dust with small fractions is transported though the structure of empty cavities in the filling filtering medium with the adsorbent of the type of *СКТ-3* with the granule's diameter of 2,5 *mm* in an air filter [4].
3) The coal dust with small fractions is captured at the surface layer in the filling filtering medium with the spherical granules with the diameter of 2 *mm* and below, creating a barrier for the aerosol stream, in an air filter (air filter captures around 80 % of the coal dust mixed with the air at the surface layer in the filling filtering medium, which can not be penetrated by the aerosol).

The absorbers of the model of *АУ-1500* in air filters at *NPP* are sustainable in relation to the critical accumulation of coal dust masses with the small dust fractions at surface layer due to the use of cylindrical filtering granules with the volume, which is close to the volume of spherical filtering granules with the diameter of 2,5 *mm*. Let us denote that the magnitude of initial air resistance by the air filters, filed with the spherical granules with the diameter of 2 *mm* (without the coal dust fraction) is in 1,5 times higher in comparison with the initial air resistance by air filters, filed with the spherical granules with the diameter of 3 *mm*. The coal dust can not penetrate though the barrier, created by this adsorbent, resulting in the malfunction by the absorber. Therefore, in the case of the air filter with the absorber of the model of *АУ-1500*, it is not recommended to fill it fully with the filtering medium, consisting of the adsorbent granules with the diameter of 2 *mm* and below.

The completed research demonstrates that the selection of right dimensions of filtering granules in the air filters at *NPP* has to be done with the consideration of existing critical limitations. The effective dimension of filtering granules must be in an agreement with the following criteria: *2 mm < d < 3 mm*. In the case of filtering granules with the smaller diameters of 2 *mm* and below, the operation resource by an air filer will be significantly limited. In the case of filtering granules with the bigger diameters of 3 *mm* and above, the absorption property by air filter will not be optimal. It is confirmed that, in the filling filtering mediums with the spherical granules, the appearing distributions of accumulated coal dust masses have some maximums of density, which are similar to the maximums, observed in the air filters with cylindrical granules [3, 4, 9].

This research paper was published in the Problems of Atomic Science and Technology (*VANT*) [10].